\documentclass{appolb}
\usepackage{graphicx}
\usepackage{xcolor}

\newcommand{\be}{\begin{equation}}
\newcommand{\ee}{\end{equation}}
\newcommand{\beq}{\begin{eqnarray}}
\newcommand{\eeq}{\end{eqnarray}}
\newcommand{\ba}{\begin{array}}
\newcommand{\ea}{\end{array}}

\begin{document}

\title{20 Years of Light Pentaquark Searches\footnote{Submitted to Acta Physica Polonica B special volume dedicated to
Dmitry Diakonov, Victor Petrov and Maxim Polyakov.}}
\author{Moskov Amaryan
\address{Old Dominion University, Norfolk, VA 23529, USA.}}

\maketitle

\begin{abstract}
In this paper, I pay tribute to my exceptional colleagues and friends Dmitri Diakonov, Victor Petrov, and Maxim Polyakov by examining the experimental progress and current status of the searches of the $\Theta^+$ pentaquark from its inception to the present.
\end{abstract}

\section {Introduction}
As this is my contribution to the memorial volume, 
I would like to begin by sharing some personal reflections.
My collaboration with Maxim Polyakov began during DIS99 Conference in Zeuthen, Germany in 1999 where I was presenting experimental data from the 
HERMES experiment at DESY, Hamburg, Germany. It was during this time that we established a friendly relationship, and I invited Maxim to give a seminar at DESY, which he graciously accepted. From that point onward, our collaboration began.

In year 2000 I obtained first results of the single beam-spin asymmetry in electroproduction of a single photon via interference of Deeply Virtual Compton scattering with a similar final state of Bethe-Heitler process.
 It was first published as a 
conference proceedings ~\cite{Amarian:2000vx} of the talk presented in 2000   at Spin2000 conference in Osaka, 
Japan, then in 2001 published in   Physical Review Letters~\cite{HERMES:2001bob}.
Now it is dubbed as DVCS and is well known to the community of DESY, CERN, JLab, and  constitutes
one of the major parts of the physics program of newly proposed Electron Ion Collider under construction 
  at Brookhaven National Laboratory, USA. Later Maxim made significant contributions 
 to observe the $D$-term for the General Parton Distributions and measurement of the pressure inside the proton. 
All this is well known and I would not like to further bother the reader with the details of these avenues.

In 2002, an international workshop was held in Santorini, Greece. Along with Maxim, Mitya Diakonov and Vitya Petrov participated in it. This is where I met them for the first time and later we collaborated closely. In this meeting
 there was no word mentioned to me about their seminal work on pentaquarks ~\cite{Diakonov:1997mm}.

In 2003 I was invited to the DIS2003 in Sankt-Petersburg, where for the first time 
I heard about the $\Theta^+$ exotic baryon made of $uudd\bar s$ quarks. I promised Diakonov, Petrov and Polyakov 
to look at $pK_S$ decay channel of the $\Theta^+$. And apparently we observed a peak in the invariant mass of $pK_{\rm S}$ 
from the HERMES data and after  a long review and discussions it was finally published ~\cite{hermes}. 
The results obtained by the HERMES collaboration initiated also other HERA experiments  to look for $\Theta^+$.
ZEUS 
Collaboration reported the observation of the $\Theta^+$ in their data ~\cite{zeus}, while H1 and HERAB didn't. 
These outcomes  created  heated debates in the DESY community, 
there was  an extraordinary seminar organized,
with all four experiments at HERA, 
where I gave an opening experimental talk and Maxim gave a theory presentation. 
Things were not conclusive as none of the HERA experiments was really designed for the hadron spectroscopy. 

In 2003 the SPRING-8 published their first observation of the $\Theta^+$ in a photoproduction on a carbon target 
 ~\cite{LEPS:2003wug}
  and independently  in the same year there was a paper by the DIANA experiment on a bubble 
  chamber with observation of the $\Theta^+$ in the $K+n\to K^0p$ reaction on the Xe target ~\cite{DIANA:2003uet}.
Consequently, in the same year the CLAS Collaboration at Jefferson Laboratory (JLab) claimed observation of the $\Theta^+$ in photoproduction experiments on a deuteron as well as on a hydrogen targets.

In the two following years there were hundreds of papers published by theorists on the topic of exotic baryons.
I may have not listed all experiments, for the detailed review I refer to \cite {hicks} and \cite{history} reviews.
All this created high excitement in the community until the CLAS Collaboration remeasured their previous 
channels with high statistics and didn't confirm their reported claims of observation of $\Theta^+$. 
This was announced on the first day of the APS meeting in Tampa, FL, in 2005 where I was invited to 
give a talk on pentaquarks two days later. 

One can imagine the level of skepticism created by the CLAS announcement. 
Anyway, I want just to remind the reader that, as Carl Sagan once said:
"Absence of evidence is not evidence of absence". I said something similar in my talk, 
however, as English proverb says: "too many cooks spoil the broth". 

Coming back to Norfolk, where I was already hired as a professor of physics at Old Dominion University 
and doing research in CLAS at JLab, I decided to look at the Quantum Mechanical interference between 
the $\phi$ and $\Theta^+$ produced with the same final state, i.e. $\gamma + p \to p \phi \to p K_{\rm S} K_{\rm L}$,
which in more detail is described in Sect.~\ref{sec:thexp}.

\bigskip

In a meantime I was thinking how to create two body reaction to answer the question of existence or non-existence 
of the $\Theta^+$ and came to the idea of creating secondary beam of neutral kaons and submitted a letter of intent to the JLab Program Advisory Committee (PAC) in 2015, which is discussed in Sect.~\ref{sec:am}. Unfortunately all my theory colleagues, 
Dmitri Diakonov, Victor Petrov and Maxim Polyakov by now passed away 
and will not see the future results.

\section{Theory and Experiment}
\label{sec:thexp}

\begin{figure}[htb!]
\centering
    \includegraphics[width=9cm]{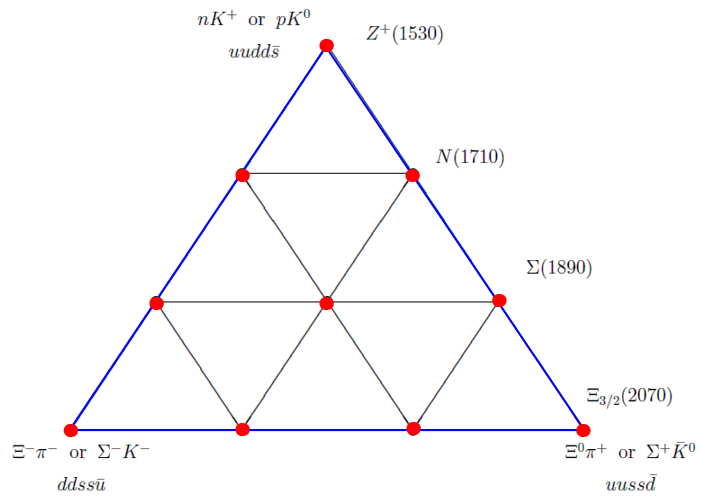} 
\caption{
The suggested anti-decuplet of baryons~\cite{Diakonov:1997mm}. The corners of this ($T_3$, $Y$) diagram are exotic.  The original name of the pentaquark lying at the apex of $\overline{10}$ was $Z^+$, then following Diakonov's suggestion, it was called $\Theta^+$.} 
\label{fig:anti10}  
\end{figure}

In his paper, classifying all existing baryons into octet and decuplet of SU(3) symmetry Gell-Mann predicted existence 
of multiquark states different from the regular 3-quark states for baryons, i.e 5-quarks in particular \cite {Gell-Mann:1964ewy}. 
However, multiple attempts over the decades didn't prove the existence of such configurations. 
The breakthrough happened after the publication of the Diakonov, Petrov and Polyakov ~\cite{Diakonov:1997mm}, 
in which authors predicted the existence of the particle in the apex of anti-decuplet, so-called $Z$-baryon, 
later dubbed as $\Theta^+$ with the mass of 1530~MeV and relative narrow width on the order of 15~MeV decaying 
either to $K^+n$ or $K^0p$. Other members of anti-decuplet were also predicted with exotics on the corners 
of the diagram, see Fig.~\ref{fig:anti10}.

There was a period of a long silence until Spring-8 ~\cite{LEPS:2003wug} and Diana experiment \cite{DIANA:2003uet} announced observation of the $\Theta^+$. Then many experimental facilities tried to observe it, see a review papers in  ~\cite{hicks} and ~\cite{history}.

Without going into the details of the Quark Soliton Model, I should mention that hundreds of papers were published in 
a short period of time until 2005 APS meeting in Tampa, Florida. There I was invited to give a talk on pentaquarks, 
but on the first day of that meeting the CLAS Collaboration
announced non-observation of $\Theta^+$ in a new set of high statistics experiments and obviously my talk looked 
like a "shot in the air". 
So since then, the situation has changed and the community has come to the conclusion that pentaquarks do not exist 
at all and we should forget about them.

Just then I started thinking 
how to try to observe $\Theta^+$ after all, if it exists.
 For that we, Maxim, Mitya and myself,  started to analyze theoretically  the interference between two 
 reactions $\gamma + p \to p \phi \to pK_{\rm S} K_{\rm L}$ and $\gamma + p \to K_{\rm S} \Theta^+ \to p 
 K_{\rm S} K_{\rm L}$ \cite{interf}.
 According to Quantum Mechanics these two processes should interfere as initial and final states of two above reactions 
 are the same. 
 
 We performed analysis of the CLAS data looking at $pK_{\rm L}$ distribution for the events selected 
under the $\phi$ peak, which was extremely clean. As a result we observed a peak in the invariant mass of $pK_{\rm L}$ 
which couldn't be explained based on the $\phi$ production alone with extensive Monte Carlo simulations. The CLAS 
Collaboration appointed a few review committees and we even created a web page where every member of the CLAS 
Collaboration could ask questions. We produced hundreds plots and answered all kinds of questions, however the CLAS Collaboration was  still reluctant to approve our analysis for the publication. 
 
\begin{figure}[htb!]
\centering
{
    \includegraphics[width=9cm]{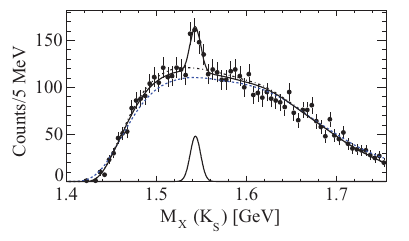} 
}
\caption {
Missing mass of $K_S$ with a cut -$t_{\Theta} < 0.45$~GeV$^2$. The dashed line is the result of a $\phi$ Monte Carlo
simulation, the dashed-dotted line is a modified Monte Carlo
distribution, and the solid line is the result of a fit with a modified
Monte Carlo distribution plus a Gaussian function.} 
\label{fig:intexp} 
\end{figure}

 The problem of three-body final state is well known, there are many overlapping resonances in the combinatorial 
 combinations and they may burry existent particle. The interference is one way to avoid such an overlap. 
 After many years of debates a small group of enthusiasts decided in 2012 to publish these results and we 
 posted the manuscript in arXiv and  
 submitted  the paper to  the Physical Review C, which in a short time was published~\cite{interexp}. This was a very long story.
 In  Fig.~\ref{fig:intexp} we show that for events under the $\phi$ peak from the fit we observe a resonance in the missing mass of $K_{\rm S}$ with the following parameters: the peak in the missing mass $M_X(K_S)$ =~1.543~$\pm 0.002$~GeV with a Gaussian width $\sigma~=~0.006$~GeV and statistical significance of 5.3$\sigma$.

Subsequently, the CLAS Collaboration published a comment paper ~\cite{Anghinolfi:2012np}, arguing that authors 
of~\cite{interexp}
used a cut on the $t$ Mandelstam variable,  which influenced the $\Theta^+$ peak. The fact that the mechanism 
of the $\phi$ or $\Theta^+$ production can change depending on the range of the $t$-Mandelstam was not accepted and was criticised
by the authors of Ref.~\cite{Anghinolfi:2012np}, 
although the subsequent paper on the $\phi$ production, where we showed that at least the mechanism of the $\phi$ production changes depending on a $t$-range ~\cite{phi}, was signed by the entire CLAS collaboration and nobody payed attention that it dismissed the counter argument in the comment paper~\cite{Anghinolfi:2012np}.

\section {Aftermaths}
\label{sec:am}

After all these twists and turns a question remains how to perform a two-body experiment with $\Theta^+$ 
produced in a formation reaction.
For that purpose one needs to search for the $\Theta^+$ in the reaction $K^0 + p \to K^+ n$.
How to make a beam of neutral kaons with a high intensity to make this possible?  As mentioned in
the introduction the letter of intent was submitted to the JLab Program Advisory Committee (PAC) in 2015.
After a few years 
the proposal  endorsed by 160 physicists from 19 countries was finally approved
in 2020 by the PAC48 for 
the secondary beam of $K_{\rm L}$ to run in Hall D of JLab for 200 days of a beam time ~\cite{KLF:2020gai}.
Since it overlaps with already approved programs in Hall D, the experiment may be scheduled to start in 2028. 
We are now working on a realization of this K-long Facility (KLF) project.

\begin{figure}[htb!]
\centering
    \includegraphics[width=8cm]{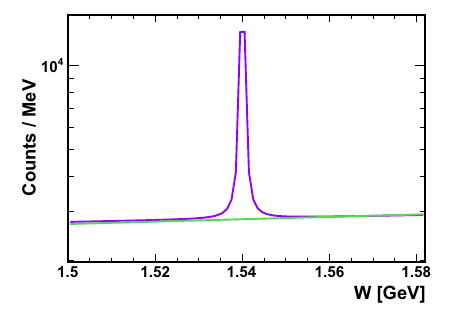} 
\caption {
Expected number of events in reaction $K_{\rm L} p \to K^+n$ as a function of W for 100 days of running on hydrogen target at GlueX setup in Hall D at JLab (for details see Ref.~~\cite{mpla}).}
\label{fig:klf} 
\end{figure}

A long story, but this is what it is. I should mention that as the pentaquark {\em per se} is a very sensitive topic, 
the proposal was written for the hadron spectroscopy without mentioning the $\Theta^+$. However, in 2024 we 
turned our attention to the reaction $K_{\rm L} + p \to K^+ n$ and we published a paper ~\cite{mpla}, where it is shown 
that if $\Theta^+$ does exist then thousands of them will be observed in 100 days of running.
As one can see from the Fig.~\ref{fig:klf} we should observe many thousands of $\Theta^+$  with a very narrow ~1-2~MeV experimental width. Otherwise, if it is not observed then one should forget it and burry the $\Theta^+$ under the stone forever, 
as the sensitivity of this reaction exceeds the level of any reasonable doubt.

\section{Acknowledgements}

I am indepted to editors of the Acta Physica Polonica providing me the opportunity to contribute to the memorial volume dedicated to my great colleagues Dmitry Diakonov, Victor Petrov and Maxim Polyakov.
This work was supportet by the U. S. Department of Energy, Office of Science, Office of Nuclear Physics, under
Award No. DE{FG02-96ER40960.


\begin{thebibliography}{99}

\bibitem{Diakonov:1997mm}
   D.~Diakonov, V.~Petrov, and M.~V.~Polyakov,
   ``Exotic anti-decuplet of baryons: Prediction from chiral solitons,''
   Z.\ Phys.\ A\ \textbf{359}, 305 (1997).

\bibitem{Amarian:2000vx}
M.~Amarian [HERMES],
``Deeply virtual Compton scattering at HERMES,''
AIP Conf. Proc. \textbf{570}, no.1, 428 (2001)
doi:10.1063/1.1384095

\bibitem{HERMES:2001bob}
A.~Airapetian \textit{et al.} [HERMES],
``Measurement of the beam spin azimuthal asymmetry associated with deeply virtual Compton scattering,''
Phys. Rev. Lett. \textbf{87}, 182001 (2001)
doi:10.1103/PhysRevLett.87.182001
[arXiv:hep-ex/0106068 [hep-ex]].

\bibitem{hermes} HERMES Collaboration, A.~Airapetian et al.,
"Evidence for a narrow $|S|$=1 baryon state at a mass of 1528 MeV in quasi-real photoproduction,"
 ~Phys. \ Lett. \ \textbf {B~ 585} (2004).

   \bibitem {zeus} ZEUS Collaboration, S.~Chekanov et al.,
 "Evidence for a narrow baryonic state decaying to $K^{0}_{S}p$ and $K^{0}_{S}\bar {p}$ in deep inelastic scattering at HERA,"
 ~Phys. \ Lett.\ \textbf {B~ 591} (2004).

 \bibitem{LEPS:2003wug}
   T.~Nakano \textit{et al.} [LEPS Collaboration],
   ``Evidence for a narrow S = +1 baryon resonance in photoproduction from the neutron,''
   Phys.\ Rev.\ Lett.\ \textbf{91}, 012002 (2003).

\bibitem{DIANA:2003uet}
   V.~V.~Barmin \textit{et al.} [DIANA Collaboration],
   ``Observation of a baryon resonance with positive strangeness in $K^+$ collisions with Xe nuclei,''
   Yad.\ Fiz.\ \textbf{66}, 1763 (2003) 
   [Phys.\ Atom.\ Nucl.\ \textbf{66}, 1715 (2003)].

\bibitem{Gell-Mann:1964ewy}
   M.~Gell-Mann,
   ``A schematic model of baryons and mesons,''
   Phys.\ Lett.\ \textbf{8}, 214 (1964).

\bibitem {hicks} Kenneth ~ H. ~ Hicks "On the conundrum of the pentaquark," ~
Eur. \ Phys. \ J. \ \textbf{H~37} (2012).

\bibitem {history} M.~Amaryan "History and geography of light pentaquark searches: challenges and pitfalls," ~Eur. \ Phys. \ J. \ \textbf{Plus~137} (2022).



\bibitem {interf} M.~Amaryan, D.~~Diakonov and M.~Polyakov "Exotic $\Theta^+$ baryon from interference," ~ Phys. \ Rev. \ \textbf{D~78} (2008).


\bibitem {interexp} M.~J.~Amaryan et al.,  "Observation of a narrow structure in $p (\gamma, K_S)X$
via interference with $\phi$-meson production," ~ Phys. \ Rev. \ \textbf {C~85} (2012).




\bibitem{Anghinolfi:2012np}
M.~Anghinolfi, J.~Ball, N.~A.~Baltzell, M.~Battaglieri, I.~Bedlinskiy, M.~Bellis, A.~S.~Biselli, C.~Bookwalter, S.~Boiarinov and P.~Bosted, \textit{et al.}
"Comment on 'Observation of a narrow structure in $p(\gamma, K_S) X$ via interference with $\phi$-meson production',''
Phys. \ Rev. \ C \textbf{86}, 069801 (2012)
doi:10.1103/PhysRevC.86.069801

\bibitem {phi} CLAS Collaboration, H.~Seraydaryan et al., "$\phi$-meson photoproduction on hydrogen in the neutral decay mode," ~ Phys. \ Rev. \ \textbf{C~89} (2014).

\bibitem{KLF:2020gai}
M.~Amaryan \textit{et al.} [KLF],
"Strange Hadron Spectroscopy with Secondary KL Beam in Hall D,''
[arXiv:2008.08215 [nucl-ex]].

\bibitem {mpla} 
Moskov J. Amaryan, Shu Hirama, Daisuke Jido, and Igor I. Strakovsky
"Search for $\Theta^+$ in $K_L p \to K^+n$ reaction in KLF at JLab," ~ 
Mod. \ Phys. \ Lett.\ \textbf{A~39}~(2024).
   





\end{thebibliography}
\end{document}